# Study of space charge in the ICARUS T600 detector*

## The ICARUS Collaboration

M. Antonello[a], B. Baibussinov[b], V. Bellini[c], F. Boffelli[d], M. Bonesini[e], A. Bubak[f], S. Centro[b], K. Cieslik[g], A. G. Cocco[h], A. Dabrowska[g], A. Dermenev[i], A. Falcone[e], C. Farnese[b], A. Fava[k], A. Ferrari[j], D. Gibin[b], S. Gninenko[i], A. Guglielmi[b], M. Haranczyk[g], J. Holeczek[f], M. Kirsanov[i], J. Kisiel[f], I. Kochanek[l], J. Lagoda[m], A. Menegolli[d], G. Meng[b], C. Montanari[d,l], C. Petta[c], F. Pietropaolo[b,j], P. Picchi[d,†], A. Rappoldi[d], G.L. Raselli[d], M. Rossella[d], C. Rubbia[j,l,n,1], P. Sala[a,j], A. Scaramelli[d,j], F. Sergiampietri[j], M. Spanu[d,2], M. Szarska[g], M. Torti[e,3], F. Tortorici[c], F. Varanini[b], S. Ventura[b], C. Vignoli[l], H. Wang[o], X. Yang[o], A. Zalewska[g], A. Zani[a].

[a] *INFN, Milano, Italy*

[b] *Dipartimento di Fisica e Astronomia "G. Galilei", Università di Padova and INFN, Padova, Italy*

[c] *Dipartimento di Fisica e Astronomia, Università di Catania and INFN, Catania, Italy*

[d] *Dipartimento di Fisica, Università di Pavia and INFN, Pavia, Italy*

[e] *Dipartimento di Fisica "G. Occhialini", Università di Milano Bicocca and INFN Milano Bicocca, Italy*

[f] *Institute of Physics, University of Silesia, Katowice, Poland*

[g] *H. Niewodniczanski Institute of Nuclear Physics, Polish Academy of Science, Krakow, Poland*

[h] *Dipartimento di Scienze Fisiche Università Federico II di Napoli and INFN, Napoli, Italy*

[i] *INR RAS, Moscow, Russia*

[j] *CERN, Geneva, Switzerland*

[k] *Fermi National Accelerator Laboratory, Batavia (IL), USA*

[l] *INFN, Laboratori Nazionali del Gran Sasso, Assergi, Italy*

[m] *National Centre for Nuclear Research, Otwock/Swierk, Poland*

[n] *GSSI, L'Aquila, Italy*

[o] *Department of Physics and Astronomy, UCLA, Los Angeles, USA*

*E-mail*: marta.torti@mib.infn.it

---

[1] Spokesperson

[2] Now Brookhaven National Laboratory (USA)

[†] Deceased

[3] *Corresponding author*

* Dedicated to the memory of Pio Picchi.

ABSTRACT: The accumulation of positive ions, produced by ionizing particles crossing Liquid Argon Time Projection Chambers (LAr-TPCs), may generate distortions of the electric drift field affecting the track reconstruction of the ionizing events. These effects could become relevant for large LAr-TPCs operating at surface or at shallow depth, where the detectors are exposed to a copious flux of cosmic rays. A detailed study of such possible field distortions in the ICARUS T600 LAr-TPC has been performed analyzing a sample of cosmic muon tracks recorded with one T600 module operated at surface in 2001. The maximum track distortion turns out to be of few mm in good agreement with the prediction by a numerical calculation. As a cross-check, the same analysis has been performed on a cosmic muon sample recorded during the ICARUS T600 run at the LNGS underground laboratory, where the cosmic ray flux was suppressed by a factor $\sim 10^6$ by 3400 m water equivalent shielding. No appreciable distortion has been observed, confirming that the effects measured on surface are actually due to ion space charge.

## Contents

## 1. Introduction

The Liquid Argon Time Projection Chamber (LAr-TPC) detection technique proposed in 1977 [1] as a modern large electronic "bubble chamber" has been taken to full maturity with the large LAr mass ICARUS T600 detector, successfully operated in 2010-2013 at the LNGS underground laboratories exposed to the CNGS beam and cosmic rays [2,3].

The ICARUS T600 detector consists of two identical modules filled with 760 t of ultra-pure liquid argon, each one housing two 1.5 m drift length TPC chambers separated by a central common cathode. A cross section of one ICARUS TPC is shown in Fig. 1. A 500 V/cm uniform electric field allows for drifting without distortions the ionization electrons produced by charged particles along their path to three parallel readout wire planes facing the drift volume and oriented at 0° and ±60° with respect to the horizontal direction. About 54000 wires in total with 3 mm pitch and plane spacing are deployed. Induced signals in the first two wire planes and the electron charge signals on the last (Collection) plane allow for measuring three independent event projections of any ionizing event with a ~1 mm space resolution. A Photo-Multiplier Tube (PMT) system, installed behind the wire planes to detect the scintillation light emitted by charged particles, is used for trigger and timing purposes [4]. Measurement of the absolute time of an ionizing event combined with the 1.55 mm/μs electron drift velocity [5] provides the position of the track along the drift coordinate. Moreover, the charge signal detected in the Collection view, which is proportional to the deposited energy, allows for the calorimetric measurement of the particle energy. The absorption of the drifting electrons by electronegative impurities is minimized by continuously filtering both liquid and gaseous argon [3,6]. Free electron lifetime in excess of $\tau_e = 7$ ms has been routinely reached in ICARUS at LNGS, corresponding to an impurity concentration < 50 $O_2$ ppt. A detailed description of the ICARUS detector can be found elsewhere [7].

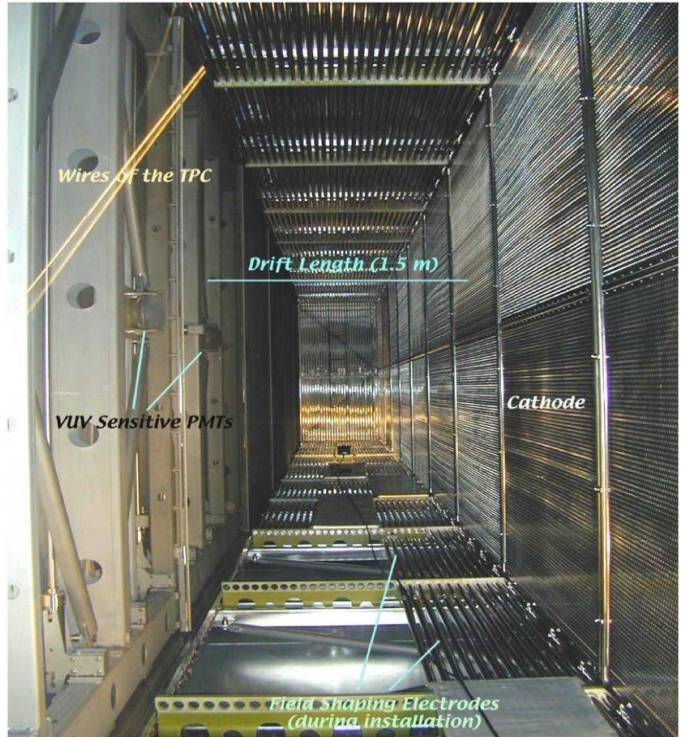

**Figure 1.** Cross section of ICARUS detector TPC during the assembling in Pavia, showing the field shaping electrodes, the cathode, the anodic wire planes and the PMT system.

ICARUS T600 will be operated at shallow depth at the Fermi National Accelerator Laboratory in the USA to search for sterile neutrinos at the Booster beam within the Short Baseline Neutrino (SBN) program [8].

In these operating conditions, space charge effects, i.e. the accumulation in the drift region of positive argon ions ($Ar_2^+$) [9,10] produced by the cosmic ray flux crossing the detector, could generate significant electric field distortions affecting the event reconstruction. In fact, due to their mobility ($\mu_i \sim 10^{-3}$ cm$^2$V$^{-1}$s$^{-1}$) [9] much smaller than the free electrons one ($\mu_e \sim 500$ cm$^2$V$^{-1}$s$^{-1}$) [11], positive ions survive in the drift region of the TPC for several minutes before being neutralized on the cathode or on the field shaping electrodes. This topic has been recently addressed by the MicroBooNE Collaboration, using a UV Laser System to measure possible space charge effects [12].

A study of space charge effects in the T600 has been performed analyzing cosmic ray induced events, collected on surface with several trigger configurations during a first technical run with one of the two ICARUS T600 modules in Pavia (Italy) [7]. The results have been compared to the ones obtained with a similar analysis carried out on a sample of cosmic muons collected during the underground operation at LNGS laboratory, where negligible space charge effects were expected due to the extremely reduced cosmic ray flux.

## 2. Modeling of space charge effects in ICARUS T600 LAr-TPC

Electrons and positive ions, created by ionizing particles in a liquid Argon TPC, drift along the same electric field lines toward the anodic wire plane and the cathode, respectively. However, due to their reduced mobility, at E = 500 V/cm argon ions drift with a velocity $v_i = \mu E \sim 5 \cdot 10^{-6}$ mm/µs, more than five orders of magnitude lower than the electron velocity, $v_e$ = 1.55 mm/µs [5]. This means that argon ions may take up to ~300 s to drift the entire ICARUS T600 anode to cathode distance, D = 1.5 m. As a consequence, a not negligible distortion in the drift field could arise because of the accumulation of positive charge in the active detector volume due to ionizing events, mainly cosmic rays.

High energy cosmic muons (> 1 GeV) are the dominant ionizing radiation at Earth's surface with a flux of about 170 muons $m^{-2}s^{-1}$ [13]. They represent the main contribution to positive space charge accumulation in a LAr-TPC operated at shallow depth with an average injected charge estimated[4] to be J = $1.7 \cdot 10^{-10}$ C $m^{-3}s^{-1}$. An additional contribution from the electromagnetic cosmic ray radiation is expected. However this soft component is mostly absorbed in the cryostat vessel and in the first 30 cm LAr thickness surrounding the 170 $m^3$ detector active volume, hence this contribution is expected to be small with respect to the crossing muon one. This is confirmed by the small rate of electromagnetic events observed in the data collected by ICARUS during the technical run on the surface in Pavia.

The rate of positive ions injected by cosmic rays has been directly estimated by measuring the total free electron charge on Collection wires for a sample of 92 triggered events. The hit finding and reconstruction procedures developed for the ICARUS experiment [14] have been adopted. As a result, a value of J = (1.9±0.1) $\cdot 10^{-10}$ C $m^{-3}s^{-1}$ has been obtained, roughly in agreement with the expectations, after applying corrections due to the hit finding efficiency (about 85%) and argon purity. This corresponds to a total energy of (3.30±0.17) GeV deposited by cosmic rays within a full drift time window t = 0.953 ms in one ICARUS T600 TPC (85 $m^3$ active volume). The ~5%

---

[4] The conversion factor between the cosmic muon flux *F* and the injected charge density *J* in LAr is:

*J [C $m^{-3}s^{-1}$]* = *F [$m^{-2}s^{-1}$]* · *dE/dx [MeV/m]* · *R* · *e [C] / W [eV]* where *dE/dx* is the average energy loss in LAr for a mip (*~210 MeV/m*), *W* is the energy required to produce an electron-ion pair in LAr (*23.6 eV*), *R* is the electron-ion recombination rate (*~0.69 at a drift field of 500 V/cm*) and *e* is the electron charge (*1.6 · $10^{-19}$ C*).

uncertainty on J is mostly related to the fluctuations in the limited event sample with similar argon purity.

As an initial evaluation, the space charge effect distortions on ionizing tracks can be derived analytically in a simple parallel plate approximation of the TPC, where the distortions at the boundaries are neglected. The space charge $\rho^+$ due to the positive ions and the electric field E in the approximation of very large parallel planar anode and cathode are determined by the Maxwell and charge continuity equations:

$$\frac{dE}{dx} = \frac{\rho^+}{\varepsilon} \tag{2.1}$$

$$\frac{d\rho^+}{dt} + \frac{d(\rho^+ v_i)}{dx} = J \tag{2.2}$$

where $\varepsilon = 1.5$ pF/m is the liquid argon dielectric constant at 87 K, $v_i$ is the ion drift velocity and $x$ is the drift coordinate ($x = 0$ and $x = D$ define the anode and the cathode positions, respectively). Introducing the dimensionless variable α [15]:

$$\alpha = \frac{D}{E_0}\sqrt{\frac{J}{\varepsilon \mu_i}} \tag{2.3}$$

where $E_0 = V/D$ is the nominal electric field in absence of space charge, eq. (2.1) and eq. (2.2) can be solved in the stationary case to derive the electric field and the space charge as a function of the drift coordinate x:

$$E(x) = E_0\sqrt{\left(\frac{E_A}{E_0}\right)^2 + \alpha^2 \frac{x^2}{D^2}} \tag{2.4}$$

$$\rho^+(x) = \frac{J\,x}{\mu_i\,E(x)} \tag{2.5}$$

where $E_A$ denotes the field at the anode. As expected, the steady state positive ion density is approximately linearly increasing from anode to cathode.

In the ICARUS technical run in Pavia, assuming the measured J = 1.9 ·10$^{-10}$ C m$^{-3}$s$^{-1}$ and the nominal electric field $E_0$ = 500 V/cm, a ~4% maximum electric field distortion due to space charge is expected, corresponding to a maximum distortion on the electron drift velocity of about 2%[5]. This effect scales approximately as $D^2$, hence it would be more relevant in case of longer drift detectors: a ~7% variation is expected on the electric field for a 3 m drift length.

To estimate the actual space charge effects, the 3D extension of eq. (2.1) and (2.2) has been numerically solved with the COMSOL® finite element package [16] to include the boundary effects of the LAr-TPC field cage. The geometrical TPC layout, the voltage applied to the cathode and to the field shaping electrodes of ICARUS T600 have been included into the computation. The uniformity of the electric field inside the TPC drift volumes when J is null, i.e. in absence of space charge, is shown in Fig. 2.

---

[5] The electron drift velocity is proportional to $E^{1/2}$ for electric field values close to $E_0$ = 500 V/cm.

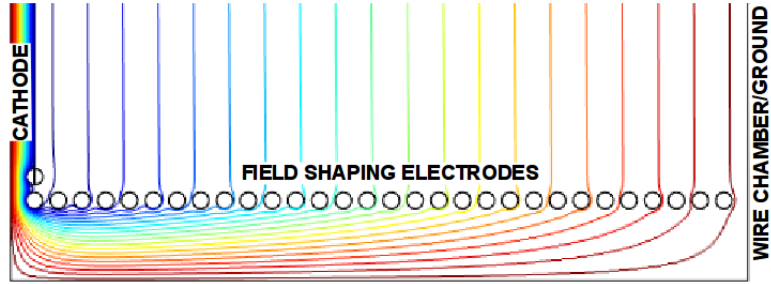

**Figure 2.** Voltage contours, in 5 kV steps, in the lower-right region of the transverse cross-section of one ICARUS T600 half-module. The blue contour corresponds to -75 kV, the red one to -5 kV.

The impact of space charge on the drift electric field is twofold with distortions along both the drift ($\Delta E_x/E_0$) and the vertical ($\Delta E_y/E_0$) directions (see Fig. 3). At the nominal drift field $E_0 = 500$ V/cm and assuming $J = 1.9 \cdot 10^{-10}$ Cm$^{-3}$s$^{-1}$ and $\mu = 0.9 \cdot 10^{-3}$ cm$^2$V$^{-1}$s$^{-1}$, the maximum calculated distortions are +4% (cathode) and -2% (anode) in x direction, and +2 % (top) and -2% (bottom) in y direction.

As shown in Fig. 3-top, the field shaping electrodes at the top and at the bottom of the TPC volume limit the distortions of the electric field $\Delta E_x/E_0$ along the drift direction. Therefore, $\Delta E_x/E_0$ in the whole volume is smaller than in the case of the analytical approximation with infinite parallel plates presented above. In particular $\Delta E_x/E_0$ is negligible in the proximity of the top and bottom electrodes, differing from the analytical approximation by ~10% in the region far from the field cage.

The distortion along the vertical direction has the effect of focusing the drift electrons towards the center of the TPC, as expected from the accumulation of a positive space charge in the bulk of the liquid argon, and it is relevant close to the field shaping electrodes (Fig. 3-bottom). Distortions along the longitudinal direction have not been considered in this study, because they are expected to be negligible due to the 18 m length of the TPC.

The analysis presented here after focuses on the local distortion of the electric field along the drift direction (Fig. 3-top). These distortions are mainly parallel to the drift direction (hence they mostly involve only a change of the electric field strength) in an inner fiducial volume far from the field cage boundaries, i.e. the region in green in Fig. 3-bottom. In this region the electric field changes can introduce a delay on the arrival time of ionization electrons at the anode, depending on the starting point of the electron cloud along the drift. The maximal delay is obtained at about 80÷90 cm of distance from the anode, as expected from the roughly linear increase of the ion density along the drift coordinate. As a result, for tracks inclined with respect to the drift direction, an apparent bending is expected in their reconstruction.

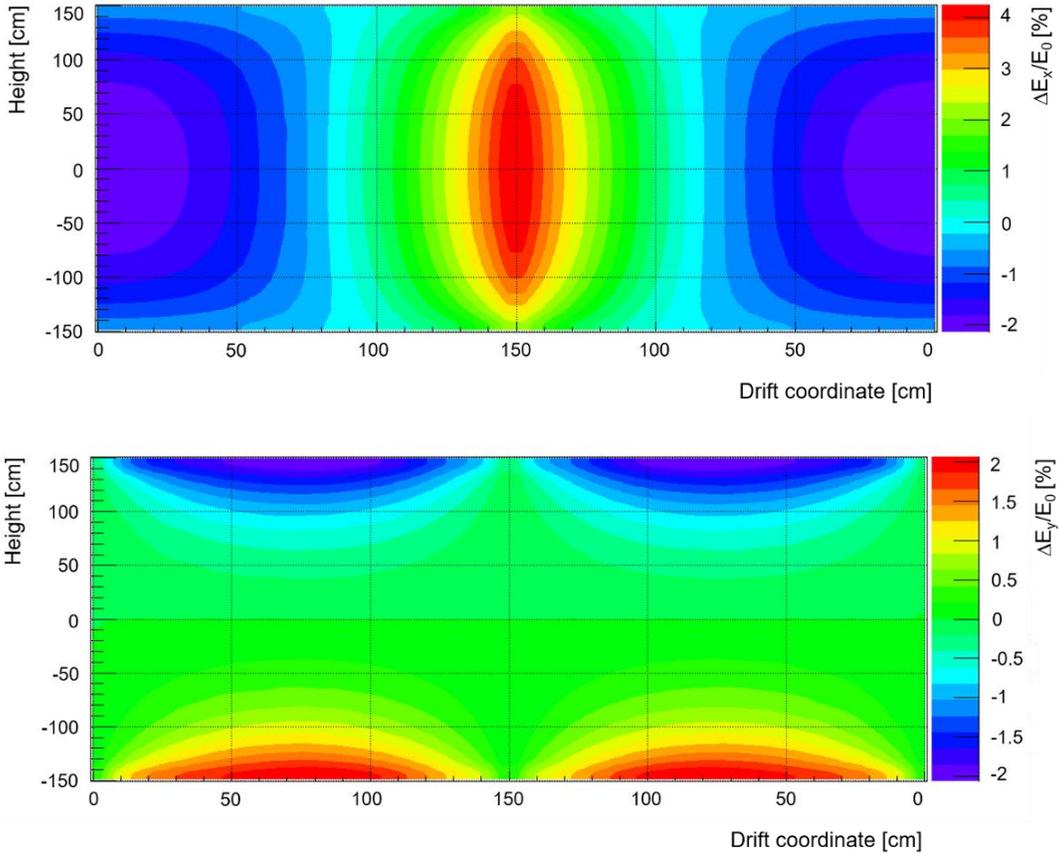

**Figure 3.** Electric field distortions $\Delta E_x/E_0$ (top) and $\Delta E_y/E_0$ (bottom) in the ICARUS TPCs as determined from an electrostatic simulation including the space charge effect from positive ions. Distortions, normalized to the nominal drift electric field magnitude ($E_0$) of 500 V/cm, are presented in a color scale in the ranges (-2%,+4%) for x and (-2%,+2%) for y. $J = 1.9 \cdot 10^{-10}$ C m$^{-3}$s$^{-1}$ and $\mu = 0.9 \cdot 10^{-3}$ cm$^2$V$^{-1}$s$^{-1}$ have been used as an input for the simulation.

Furthermore, the presence of electro-negative impurities (mainly $O_2$, $CO_2$ and $H_2O$) in the LAr can introduce a limited negative space charge through the attachment of the drifting ionization electrons before they reach the anodic plane. This effect becomes relevant when the associated electron lifetime $\tau_e$ is comparable to the maximum drift time (1 ms). In the ICARUS run in Pavia a lifetime $\tau_e = 1.7$ ms (±8%) was measured [5], hence on average ~25% of the ionization electrons were captured before reaching the wire planes. The resulting distribution of negative ions, increasing from cathode to anode, significantly impacts the overall space charge effect. Therefore, the contribution of negative ions has also been taken into account in the simulation, assuming similar mobility for negative and positive ions [10,17]. The space charge density due to negative ions injection rate then takes the following functional behavior:

$$\rho^-(x) = -\frac{J}{\mu_i E(x)}\left[(D-x) - v_d\tau_e\left(1 - e^{-\frac{D-x}{v_d\tau_e}}\right)\right] \qquad (2.6)$$

Note that the electric field distortion shown in Fig. 3 has been simulated in the ideal case of infinite drift electron lifetime as a matter of example, but for the analysis presented in the following both the positive and negative ion densities have been included in the simulation.

Finally, thermal-induced convective motions of liquid argon, with speeds similar to that of the ions, could also affect the ion density and the bending shape of the ionizing tracks. They are not included in the simulation, but their effect could emerge in the data as deviations in the bending shape of the ionizing tracks with respect to the calculated one.

## 3. Space charge effect measurements in ICARUS T600

During the forthcoming ICARUS T600 operation at FNAL in the framework of the SBN programme, the detector could be affected by space charge distortions, being operated at shallow depth. ICARUS T600 experienced similar experimental conditions during an initial test run in Pavia, where one of the two modules was operated at surface. The cosmic muons collected during this run can be used to estimate the expected charge effects before the ICARUS data taking at FNAL [18].

For this purpose, a sample of cosmic muon bundles has been selected (see Fig. 4) for the present analysis. The high energy of muons mitigates deviations from a straight line due to Multiple Coulomb Scattering. In addition, all the muons being parallel, it has been possible to analyze the tracks together to enhance coherent deviations from straight lines due to global effects, such as ion space charge. These events also contain a large number of easily identified tracks crossing the full drift region, allowing to select the tracks spanning the fiducial volume far from the field cage boundaries, where the electric field distortions are essentially along the drift direction.

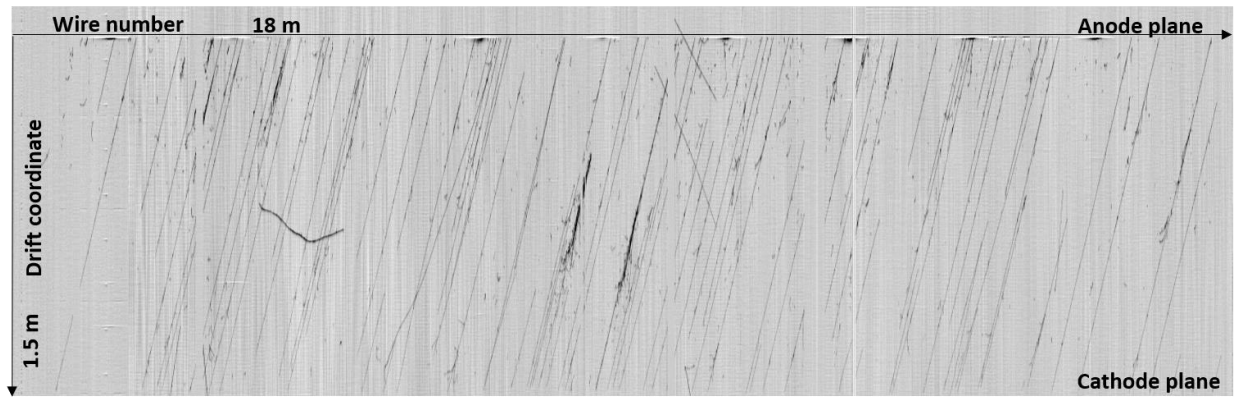

**Figure 4.** Example of a muon bundle recorded at surface with the ICARUS T600 detector. Anode and cathode are evidenced, as well as wire and drift coordinates.

*3.1 Analysis of a surface cosmic rays sample*

The whole event sample used in this analysis consists of 67 muons recorded in a data taking period when the drift electron life-time was 1.7 ms. Tracks crossing the entire drift distance have been selected by visual scanning in order to cover uniformly the whole active volume. The standard ICARUS hit finding, fitting and clustering procedures [14] have been applied, as depicted in Fig. 5, with the assumption of an ideal uniform electric drift field corresponding to a constant drift velocity throughout the full volume. Delta rays have been removed to provide a pure muon track sample. The initial and final 1.5 cm of each track have been also excluded from the muon cluster, to mitigate possible boundary non-uniformities in the proximity of the wires and of the cathode plane.

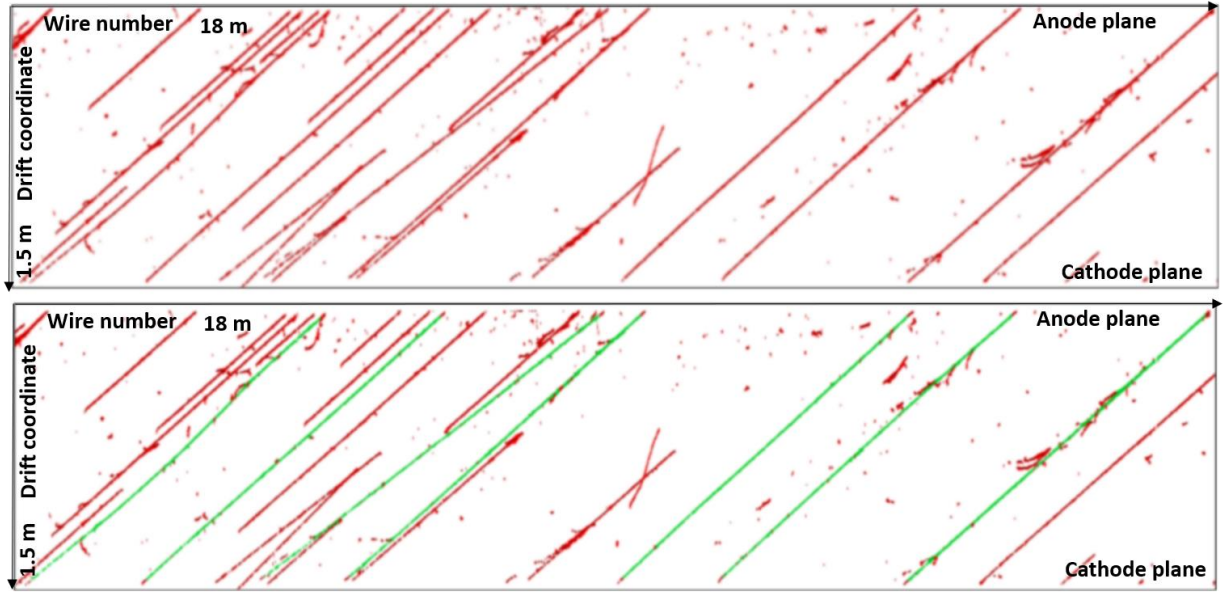

**Figure 5.** Enlarged view of a muon bundle event recorded in one chamber. Top: hits identified for all tracks are shown. Bottom: green points correspond to muon tracks that crossed both cathode and wire planes. Muon tracks do not include the first and last 1.5 cm in order to avoid boundary non-uniformities.

An apparent bending of reconstructed muon tracks could suggest the presence of ion accumulation in the liquid argon. In fact, due to the dependence of the electron drift velocity on the electric field, the ionization electron arrival time on the anode will be delayed with respect to the time observed in case of uniform electric field, as described in Sec. 2. Then, the apparent bending of muon tracks along the drift coordinate can be estimated through the time delay $\Delta T$ defined as:

$$\Delta T = T_{sc} - T_u \qquad (3.1)$$

where $T_{sc}$ is the actual electron arrival time at the anode and $T_u$ is the corresponding drift time with respect to the time expected in case of a uniform electric field (see Fig. 6 left). The time delay $\Delta T$ as a function of the drift distance is shown in Fig. 6 right, for the analytical case and with positive ions, only for sake of illustration. It has to be noted that the time delay at the cathode is expected to be small but not zero because of the presence of non linear effects due to space charge[6]. A linear fit of the initial and final 5 cm of each muon track, which are roughly insensitive to the presence of space charge, is performed to determine an undistorted ionizing track and the corresponding electron arrival times $T_u$ on the interested TPC wires. The bending parameter $\Delta T$ is calculated hit by hit for the full track and converted to a spatial deviation $\Delta X$ through:

$$\Delta X = v_e \cdot \Delta T \qquad (3.2)$$

where $v_e = 0.155$ cm/μs is the electron drift velocity for an electric field of 500 V/cm measured in ICARUS during its operation on surface [5].

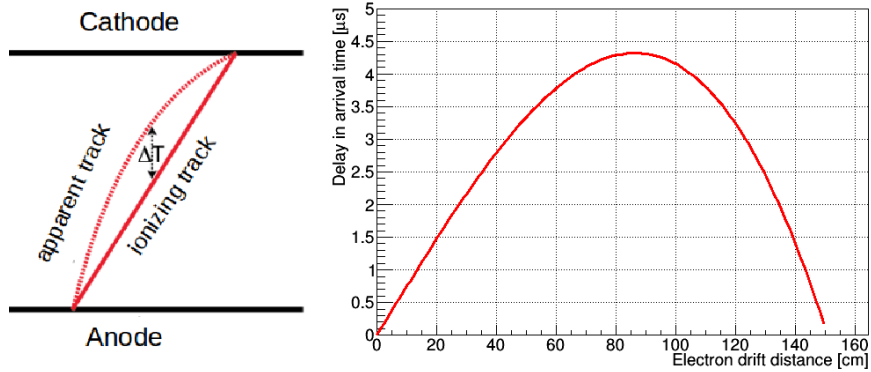

**Figure 6.** Left: schematic view of the bending parameter $\Delta T$. Right: delay in arrival time for drift electrons along the full drift, evaluated in the analytical case and with positive ions only for $J = 1.9 \cdot 10^{-10}$ C m$^{-3}$s$^{-1}$ and $\mu = 0.9 \cdot 10^{-3}$ cm$^2$V$^{-1}$s$^{-1}$. The anode position corresponds to the 0 cm in drift coordinate, while the cathode position is at 150 cm.

The predicted distortion $\Delta X$ in eq. 3.2 is obtained from the Electric field map by integrating the time delay of the electrons drifting along the field lines from the production point to the anode (Fig. 7 top). Similarly the corresponding $\Delta Y$ distortion is evaluated as the vertical displacement of the electric field lines from the production point to the anode (Fig. 7 bottom). The values $J = 1.9 \cdot 10^{-10}$ C m$^{-3}$s$^{-1}$ and $\mu = 0.9 \cdot 10^{-3}$ cm$^2$V$^{-1}$s$^{-1}$ have been used as an input to the simulations.
To minimize the bowing introduced by the vertical component $E_y$ of the electric field, tracks have been selected to be fully contained in a region where the vertical distortions ($\Delta E_y/E$) along the whole ionization electron trajectory are expected to be <0.5%. The resulting subsample of 40 tracks characterized by similar directions and affected by homogeneous electric field distortions is shown in Fig. 7 overlapped with the calculated map of the longitudinal spatial distortions along the drift (top) and vertical (bottom) directions.

---

[6] Around the nominal electric field value $E_0 = 500$ V/cm, the electron drift velocity scales approximately as $E^{\frac{1}{2}}$. In the approximation of $\varepsilon(x) = \frac{\Delta E_x}{E_x} \ll 1$, the actual electron arrival time at the anode can be evaluated as $T_{sc} = \int_0^D \frac{dX}{v(x)} = \int_0^D \frac{dX}{\mu E_0 \sqrt{1+\varepsilon(x)}} = \int_0^D \frac{dX}{\mu E_0}(1 - \frac{1}{2}\varepsilon + o(\varepsilon^2)) = T_u + \int_0^D \frac{dX}{\mu E_0} o(\varepsilon^2)$. In fact, the linear term $\int_0^D \varepsilon(x)dx = 0$ because of the boundary condition on the voltage V applied on the cathode which is fixed independently from the local electric field values in the drift volume: $V = \int_0^D E(x)dx = \int_0^D E_0(1 + \varepsilon(x))dx = \int_0^D E_0 dx + \int_0^D \varepsilon(x)dx = V + \int_0^D \varepsilon(x)dx$.

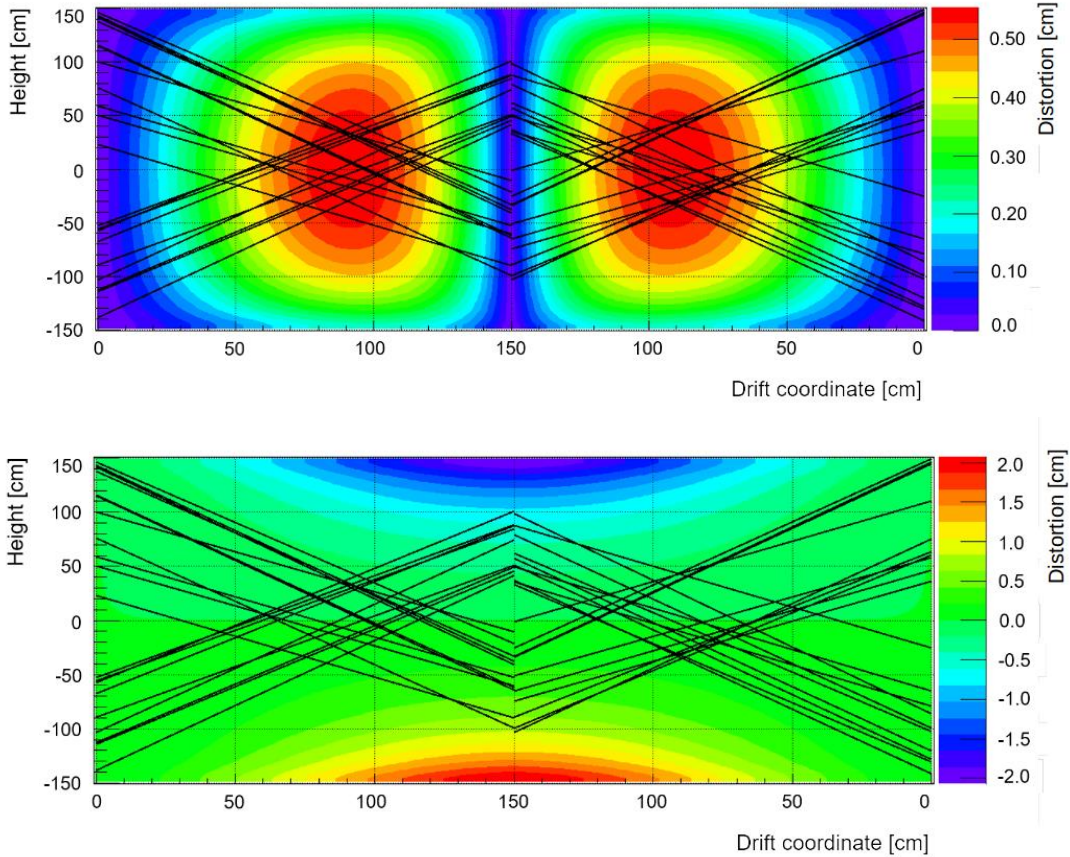

**Figure 7.** The 40 tracks selected for the analysis overlapped with the map of the longitudinal distortions along the drift (top) and vertical (bottom) directions, both expressed in cm in the color scale. Tracks crossing the regions where the transverse distortion of the electric field was larger than 0.5% have been discarded from the analysis. $J = 1.9 \cdot 10^{-10}$ C m$^{-3}$s$^{-1}$, $\mu = 0.9 \cdot 10^{-3}$ cm$^2$V$^{-1}$s$^{-1}$ and $\tau = 1.7$ ms as in the real data, corresponding to 0.18 ppb O$_2$ equivalent, have been used as an input for the simulation.

The various simulations of the electric field distortions have been used to evaluate the apparent track bending with increased accuracy, starting from the ideal case of infinite parallel plates to the real case of the selected track sample. To this purpose, the initial and final coordinates of the 40 tracks have been measured to build the corresponding straight tracks used as an input for the calculation, relying on the fact that the signal at the anode does not undergo any distortion, while at the cathode the distortions are small but not null, as visualized in Fig. 8.

In the actual ICARUS detector configuration and for the analysed event sample the maximum measured distortion ΔX differs from the ideal case of infinite parallel plates by ~40% due to a number of effects, as illustrated in Fig. 8 for $J = 1.9 \cdot 10^{-10}$ C m$^{-3}$s$^{-1}$ and $\mu = 0.9 \cdot 10^{-3}$ cm$^2$V$^{-1}$s$^{-1}$.

First of all, a ΔX reduction of ~6% is introduced by the ICARUS detector structure taking into account the geometry of the field shaping electrodes. The contribution to space charge from negative ions mitigates the maximum track distortion by ~12%, with a ~1% uncertainty coming from the 8% measurement error on the free electron lifetime. As a result, a maximum distortion ΔX = 6.7 mm, in the ideal case of infinite parallel plates, is reduced to 5.5 mm when these effects are included (see Fig. 8).

An additional systematic shift, due to the tracks crossing the detector regions where the electric field distortions are not maximal, has to be determined for the selected track sample (Fig. 7). Its value can be precisely evaluated on a track by track basis, each one reconstructed with a few mm precision. For the 40 tracks sample available for this study, a ~14% systematic reduction has been evaluated.

Finally, the linear fit used in the track reconstruction method, where the direction of each muon is set by the first and last 5 cm of the track, introduces a ~8% systematic reduction, bringing the measured maximum ΔX to 4 mm (see Fig. 8).

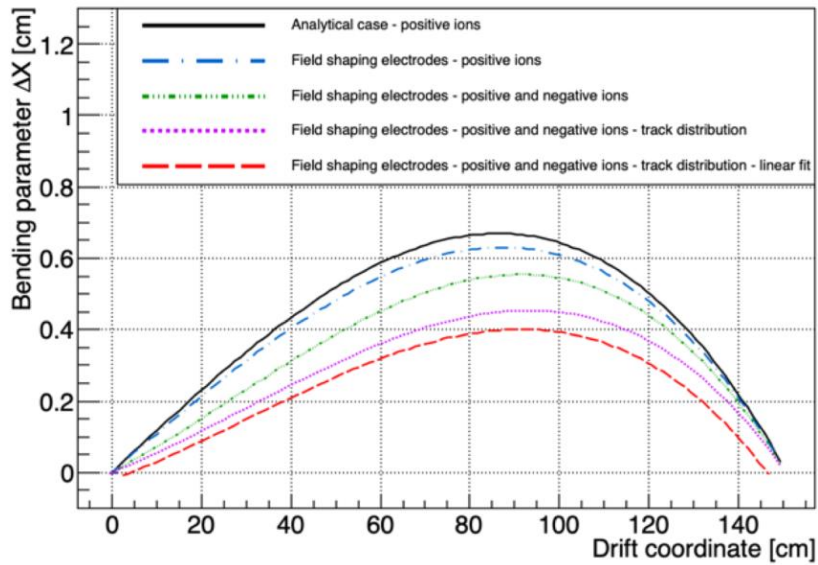

**Figure 8.** Bending parameter ΔX as a function of the drift coordinate as evaluated from simulations with J = $1.9 \cdot 10^{-10}$ C m$^{-3}$s$^{-1}$ and $\mu$ = $0.9 \cdot 10^{-3}$ cm$^2$V$^{-1}$s$^{-1}$ as input parameters: analytical formula referring to infinite parallel plates and positive ions only (black continuous line); simulation of the ICARUS actual geometry with the field shaping electrodes (blue dotted-dashed line); contribution of the negative ions (green dotted-dashed line); effect of the available sample of 40 tracks (purple dotted line); reduction due to the track fitting method (red dashed line).

*3.2 Results from surface cosmic data*

The presence of a track bending effect has been initially investigated evaluating the ΔX parameter as a function of the drift coordinate (Fig.9-left). On a track-by-track basis the Multiple Coulomb Scattering (MCS) fluctuations dominate, being the outliers associated with lower energy muons undergoing larger MCS. However, a global bending effect of up to a few millimeters clearly emerges. A more refined analysis has been performed by dividing the drift path into 15 intervals, 10 cm wide, and calculating the average ΔX with a gaussian fit inside each bin, thus making ΔX insensitive to the large fluctuations induced by MCS. The measured deviations from the straight track are shown for the two ICARUS TPCs separately in Fig. 9-right. The corresponding maximal deviation of 0.4 cm is observed at ~90 cm distance from the anode, as expected from the previous calculation. Moreover, the observed bending effect respects the symmetry of the detector and no significant differences are observed between the left and right TPCs.

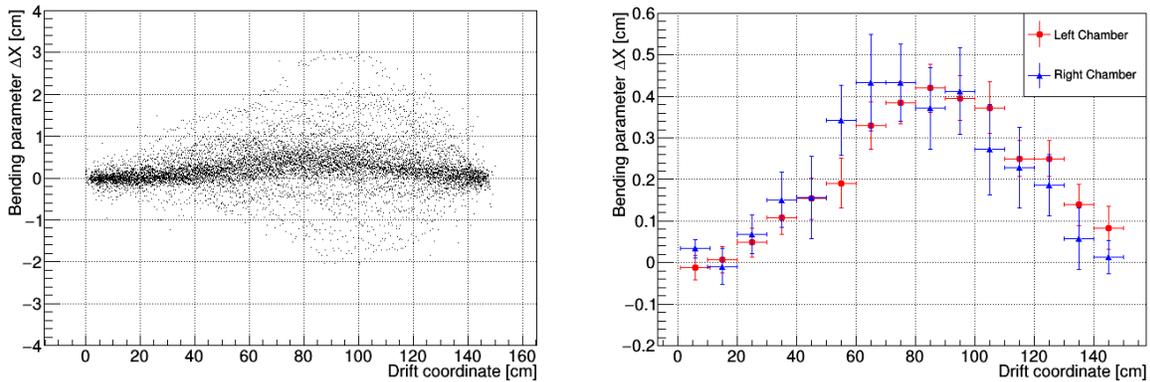

**Figure 9.** Left: scatter plot of bending parameter ΔX as a function of the drift coordinate for all selected tracks of Pavia surface sample. Right: average bending parameter ΔX as a function of the drift coordinate evaluated on fixed drift length segments (10 cm) for the two TPCs separately. The anode position corresponds to the 0 cm in drift coordinate, while the cathode position is at 150 cm.

These results have been compared with the expectations from the space charge simulation for few values of ion mobility, including the contribution of the negative ions for the measured electron lifetime $\tau_e = 1.7$ ms, see Fig. 10. A good agreement with the measured maximum track distortion of ~4 mm is found for $J = (1.9 \pm 0.1) \cdot 10^{-10}$ C m$^{-3}$s$^{-1}$ and $\mu \sim 0.9 \cdot 10^{-3}$ cm$^2$V$^{-1}$s$^{-1}$.

As already discussed in Sec. 2, eq. (2.5), the space charge distortions are proportional to the ratio $J/\mu$, hence the rough estimation of the average current density J injected in the detector by cosmic rays actually prevents to quote a precise value for the mobility parameter $\mu$.

The used space charge model well describes the general features of the measurements, thus providing a first hint that the observed effect of track bending is actually due to space charge effects.

Accounting for the systematic effects related to the event sample and fit procedure (see Sec. 3.1), a maximum $\Delta X_{max} = 5.5 \pm 0.5$ mm can be estimated at the center of the detector, far from the field shaping electrodes.

The slight shape disagreement could be attributed to residual effects not included into the simulation, such as global flows introduced by the liquid recirculation and thermal-induced convective LAr motions which have typical speed comparable to that of the ions, as well as further effects as recently suggested in [19].

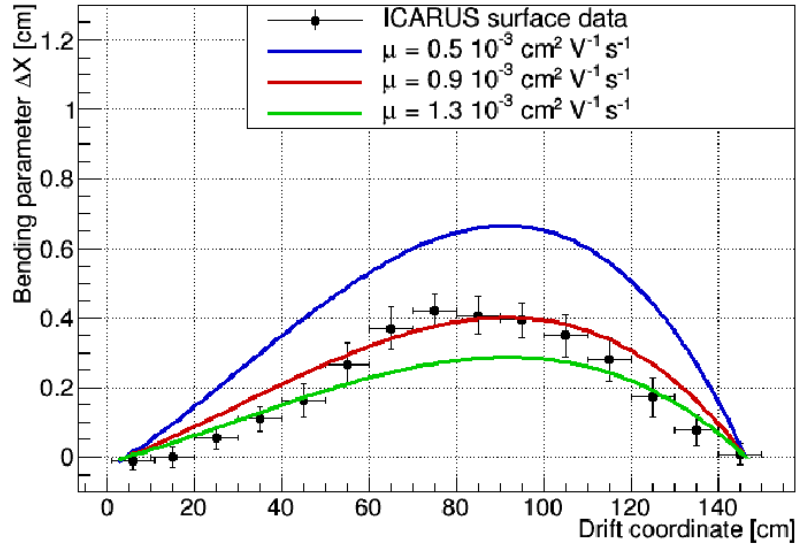

**Figure 10.** Bending parameter ΔX as a function of the drift coordinate. Simulations for three different values of the ion mobility, including also negative ions, are compared with the experimental data (black dots). The value $J = 1.9 \cdot 10^{-10}$ C m$^{-3}$s$^{-1}$ has been used as an input for the simulation.

*3.3 Results from underground cosmic data*

A further confirmation about the measured space charge effects in the Pavia dataset comes from the analysis of cosmic ray data collected by ICARUS T600 underground at LNGS, where the cosmic ray flux is largely suppressed and the space charge effects should be negligible. In underground conditions muon bundles are not as frequent as on surface, so single muon events have been used in the analysis. A sample of 81 muon tracks has been analyzed with the same selection criteria and methods described for the surface cosmic muon tracks. As expected, in the LNGS sample the result is compatible with absence of space charge effects, as shown in Fig. 11, where the Multiple Coulomb Scattering effect on the single track is dominating.

It follows that the bending effects as measured on cosmic muon tracks in ICARUS on surface can be indeed associated with the presence of space charge inside the LAr active volume.

As a final remark, the absence of distortions in the underground measurements does not allow elucidating the impact of the convective motions in the liquid argon on the distortions observed at surface. In fact, at LNGS the positive ion accumulation is negligible due to the reduced cosmic ray flux and the negative ions density is even smaller due to the high drift electron lifetime. As a consequence the space charge is not affecting the drifting electron trajectories and any liquid motion redistributing the ion density is also not measurable.

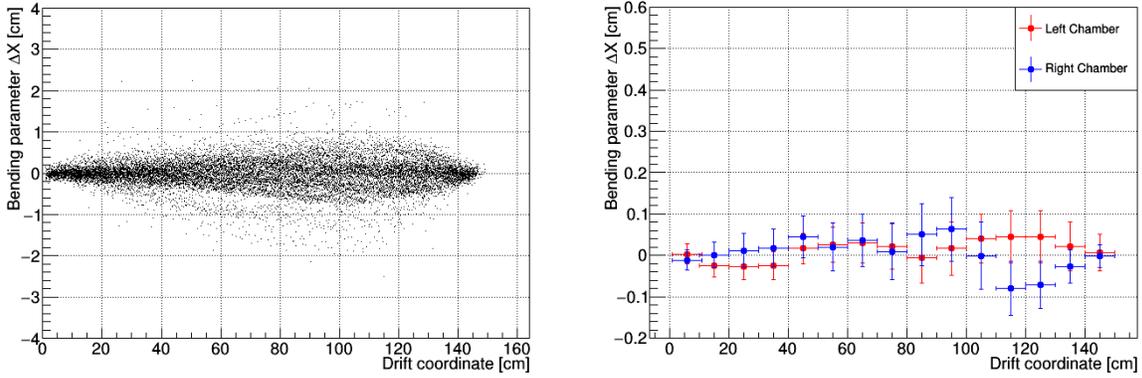

**Figure 11.** Left: scatter plot of bending parameter ΔX as a function of the drift coordinate for all tracks of LNGS underground sample. The observed fluctuations on ΔX are compatible with the Multiple Coulomb Scattering effect. Right: bending parameter ΔX as a function of the drift coordinate evaluated for the two TPCs separately.

## 4. Conclusions

The analysis performed on the ICARUS data collected at surface in Pavia demonstrates the presence of small space charge effects with a maximum track bending value $\Delta X_{max} = 5.5 \pm 0.5$ mm, when corrected for the systematics effects introduced by the adopted analysis method.
The obtained results are in good agreement with an electric field simulation that takes into account both the positive ion accumulation, mainly from cosmic ray ionization, and the negative one from ionization electron capture by electronegative impurities.
On the other hand, no track bending was observed on a cosmic muon sample collected underground by ICARUS at LNGS, confirming the distortions observed at surface are indeed due to ion space charge.
It follows that space charge effects will not be a critical issue for the forthcoming ICARUS T600 detector operations at shallow depth at Fermilab, but shielded from cosmic rays by 3 m concrete overburden. In fact, the observed track bending is expected to be of the order of the detector spatial resolution, considering the additional mitigation coming from the ~30% cosmic ray flux reduction from the overburden shielding the ICARUS detector [8].

**Acknowledgments**

The ICARUS Collaboration dedicates this work to the memory of Pio Picchi who pioneered the LAr TPC technology. The ICARUS Collaboration acknowledges the fundamental contribution of INFN to the construction and operation of the experiment. In particular, the authors are indebted to the LNGS Laboratory for the continuous support to the experiment. The Polish groups acknowledge the support of the National Science Center, Harmonia (2012/04/M/ST2/00775).

**References**

[1]  C. Rubbia, The Liquid-Argon Time Projection Chamber: A New Concept For Neutrino Detector, CERN-EP/77-08 (1977).
[2]  C. Rubbia et al. (ICARUS Collaboration), 2011 JINST **6** P07011.
[3]  M. Antonello et al. (ICARUS Collaboration), 2015 JINST **10** P12004.
[4]  M. Antonello et al. (ICARUS Collaboration), 2014 JINST **9** P08003.
[5]  S. Amoruso et al. (ICARUS Collaboration), Nucl. Instr. and Meth. **A516** (2004) 68.
[6]  M. Antonello et al. (ICARUS Collaboration), 2014 JINST **9** P12006.
[7]  S. Amerio et al. (ICARUS Collaboration), Nucl. Instr. and Meth. **A527** (2004) 329-410.
[8]  R. Acciari et al., *A Proposal for a Three Detector Short-Baseline Neutrino Oscillation Program in the Fermilab Booster Neutrino Beam*, arXiv:1503.01520.
[9]  T.H. Dey and T.J. Lewis 1968 J. Phys. D: Appl. Phys. 1 1019.
[10] R.L. Henson, Phys. Rev. 135 (1964) 1002.
[11] National Institute of Standards and Technology, Web site: http://www.nist.gov/.
[12] C. Adams et al., *A Method to Determine the Electric Field of Liquid Argon Time Projection Chambers Using a UV Laser System and its Application in MicroBooNE*, arXiv:1910.01430
[13] M. Tanabashi *et al.* (Particle Data Group), Phys. Rev. **D98**, 030001 (2018) and 2019 update.
[14] M. Antonello et al. (ICARUS Collaboration), Adv. High Energy Phys. 2013 (2013) 260820.
[15] S. Palestini et al., Nucl. Instr. and Meth. **A421** (1999) 75.
[16] COMSOL® Multiphysics, http://www.comsol.com .
[17] H.T. Davis, S.A. Rice and L. Meyer, J. Chem. Phys. **37** (1962) 947; J. Chem. Phys. **37** (1962) 2470.
[18] M. Torti on behalf of ICARUS Collaboration, EPJ Web of Conferences **126** (2016) 05013.
[19] X. Luo and F. Cavanna 2020 JINST 15 C0303.